\documentclass[12pt]{article}
\usepackage{amsmath}
\usepackage{epsf}
\usepackage{amsfonts}

\def\({\left(}
\def\){\right)}
\def\[{\left[}
\def\]{\right]}
\def\non{ \nonumber }   

\def\ctimes{\stackrel{\otimes}{,}}

\begin{document}
\rightline{LPTHE-01-61}
\vskip 2cm
\centerline{\LARGE On quantization of affine}
\centerline {\LARGE Jacobi varieties of spectral curves.} 
\vskip 2cm
\centerline{\large F.A. Smirnov\footnote[0]{Membre du CNRS}, V. Zeitlin
}
\vskip1cm
\centerline{ Laboratoire de Physique Th\'eorique et Hautes
Energies \footnote[1]{\it Laboratoire associ\'e au CNRS.}}
\centerline{ Universit\'e Pierre et Marie Curie, Tour 16, 1$^{er}$
		\'etage, 4 place Jussieu}
\centerline{75252 Paris Cedex 05, France}
\vskip2cm
\noindent
{\bf Abstract.} A quantum integrable model related to
$U_q(\widehat{sl}(N))$ is considered. A reduced model is introduced which 
allows interpretation in terms of quantized affine
Jacobi variety. Closed commutation relations
for observables of reduced model are found.

\newpage

\section{ Classical case.}
\vskip 0.5cm
\noindent

Consider classical integrable model with the $l$-operator
which is an $N\times N$ matrix depending on the spectral
parameter $z$:
\begin{align}
& l(z)=l^+(z)+ l^0(z)+zl^-(z),
\label{l}
\end{align}
$l^{\pm}(z)$ are polynomials of degree $n-1$, 
$l^0(z)$ is polynomial of degree $n$,
$l^{+}(z)$ ($l^{-}(z)$) is upper (lower)-triangular,
$l^0(z)$ is diagonal.
The classical algebra of observables $\mathcal{A}$ is
generated by the coefficients of polynomials 
giving the matrix elements of $l(z)$.
The algebra $\mathcal{A}$ a is Poisson algebra, 
Poisson structure being given by the $r$-matrix relations:
\begin{align}
&\{l(z) \ctimes l(z')\}=
\[r(z,z'),l(z)\otimes l(z')\]
\non
\end{align}
where the classical $r$-matrix is
\begin{align}
&r(z,z')=
\frac {(z+z')}{2(z-z')} \sum\limits _{i}E^{ii}\otimes E^{ii}
+\non\\ &+\frac {z}{z-z'}\sum\limits _{j>i}E^{ji}\otimes E^{ij}
+\frac {z'}{z-z'}\sum\limits _{j<i}E^{ji}\otimes E^{ij}
\non
\end{align}
with
\begin{align}
      &\begin{matrix}\  \ &\ &\ \ j &\ \end{matrix}\non\\
E^{ij}= &\begin{pmatrix} 0 &\vdots &0\non\\
                         \cdots &1 &\cdots \non\\
                         0 &\vdots &0
        \end{pmatrix}
        \begin{matrix}
 \ \non\\i\non\\ \ 
        \end{matrix}\non
\end{align}
Let us introduce the polynomials $t_k(z)$:
\begin{align}
&\text{det}(wI+l(z))=\sum\limits_{k=0}^N w^{n-k}t_k(z)
\non
\end{align}
and there coefficients $t_k^{(j)}$ defined by
$$t_k(z)=\sum\limits_{j=0}^{kn}t_k^{(j)}z^{nk-j}$$

The form of the Poisson brackets implies that all coefficients of
characteristic polynomial are in involution:
$$
\{\text{det}(wI+l(z)),\text{det}(w'I+l(z'))\}=0
$$
Moreover, part of them belongs to the center of the Poisson brackets:
these are
$$t_N^{(j)} \quad \quad \forall j$$
because of
$$\{\det(l(z)) \ctimes l(z')\} = 0$$
and
$$t^{(kn)}_k = t_k(0) \quad \quad \forall k $$
because $r(z, z')$ becomes a constant matrix for $z = 0$.

Let us fix the center. Then the remaining phase space $\mathcal{M}$ is of
dimension:
$$\text{dim}(\mathcal{M})=nN(N-1)$$ 
The number of remaining integrals
of motion (of coefficients $t_k^{(j)}$ except the central ones) is
$$\frac 1 2 nN(N-1)=\frac 1 2 \text{dim}(\mathcal{M})$$
So, the number of integrals of motion in involution is exactly one half
of the dimension of the phase space and, hence, we are dealing
with an integrable model.

However, there is one subtlety here. We consider some specific
real form of the model (without going into the details at
this point). For the real form in question the level of integrals looks
as follows:
\vskip 1cm
\hskip 6cm
\epsffile{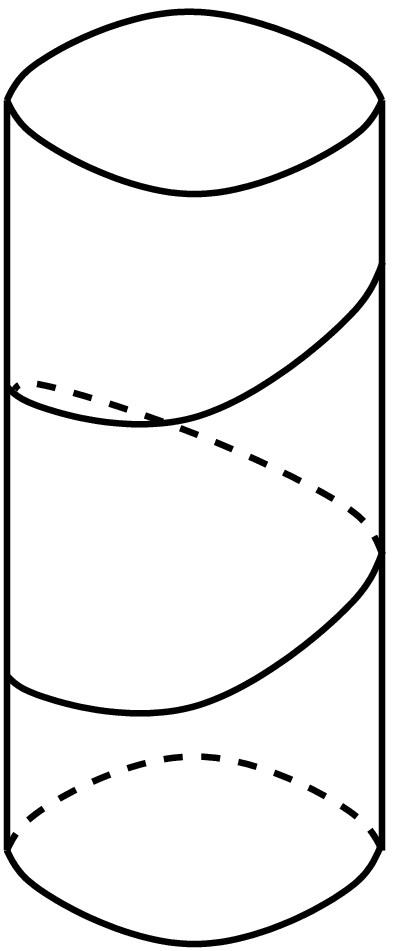}
\vskip 0.3cm
\noindent
{\it Figure 1.}
\vskip 0.2cm
\noindent

The integrals are divided into ``compact'' and ``non-compact'' ones.
The typical integral curve of Hamiltonian vector-field
corresponding to ``compact'' integral is the spiral on
{\it Figure 1} while the one corresponding to ``non-compact'' integral
is strictly vertical.
We would like to reduce the model in order that 
the ``non-compact'' integrals disappear and the integral
curves of ``compact'' ones close.
Such reduction is important for several reasons:
\newline
1. To consider average and adiabatic approximation we need compact
levels of integrals.
\newline
2. In the quantum case non-compact directions must
correspond to continuous spectrum which we would like to eliminate.
\newline
3. The integrable model in question can be viewed as lattice approximation
of CFT. Non-compact directions correspond to zero-modes. We would like
to separate them from the rest of degrees of freedom in order to
have clear identification of primary fields.
\newline
4. There is an algebra-geometric reason to eliminate the
non-compact directions on which we would like to make some detailed
comments.

It is well known that the integrable model in question is 
closely related to the algebraic geometry through the study
of the spectral curve:
$$X:\ r(w,z)\equiv\text{det}(wI+l(z))
=0$$
of genus: $g=\frac 1 2 (N-1)(Nn-2)$.
The relation is as follows. 
Consider the  Jacobi variety of the curve $X$ i.e. the complex torus:
$$J=\frac{\mathbb{C}^g}{\mathbb{Z}^g\times B\mathbb{Z}^g}$$
where $B$ is the period matrix.
Introduce corresponding 
Riemann Theta-function which satisfies:
$$
\theta (\zeta +m+B n)=
\text{exp}\ 2\pi i
\(-\textstyle{\frac{1}{2}}\ {}^t nB n -{}^t n \zeta\) \ \theta (\zeta ),
\qquad \forall m, n \in \mathbb{Z}^g
$$
The relation between integrable models and algebraic geometry is 
normally based on the following ``identity'':
\vskip .2cm
\centerline{\it Liouville torus = Real part of Jacobi variety.}
\vskip .2cm
\noindent
However, this relation only holds in the usual cases because the levels
of the ``non-compact'' integrals are usually fixed from the very beginning
in the particular integrable model under consideration, so these integrals
rarely, if ever, appear explicitly in the discussion (the problem of having
``extra'' degrees of freedom resulting from imposing some additional implicit
constraints is well-known though).

But in our case this ``identity'' cannot be correct as can be easily seen
from the comparison of dimensions: 
$$\text{dim}(\mathcal{M})=2g + 2(N-1)$$
In fact, the real part of Jacobi variety describe
only the compact part of the level of integrals and, clearly, to use the
usual algebra-geometric formulation we need to eliminate $N-1$ non compact
integrals.
Let us discuss this point in some more details.

The integrable model under consideration allows a complexification.
Complexification of the compact part of the level of
integrals should give the Jacobi variety. 
More precisely, the complexification 
gives the Jacobi variety 
from which the following divisor is cut off:
$$D=\{\zeta \in J |
\ \theta (\zeta +\rho _1)\cdots \theta (\zeta +\rho _N)=0\}$$
where $\rho _1,\cdots ,\rho _N$ are images under Abel map of
$N$ points of $X$ which project onto the point $\infty$
on $z$-plane.
In other words the observables considered as functions on the Jacobi
variety possess singularities (only) on $D$.
Generally we have:
\vskip .2cm
\centerline{\it Functions on the Level of Integrals of Motion =}
\vskip .2cm
\centerline{{\it =(Functions on $J_{\text{aff}}$) $\times$ (Sections)}}
\vskip .2cm
\noindent
where $J_{\text{aff}} \equiv J-D$ stands for affine Jacobi variety.
{\it Sections} correspond exactly to non-compact directions
of the level of integrals. In algebra-geometric language they
are given by expressions of the form
$$\frac {\theta (\zeta +\rho _i)}{\theta (\zeta +\rho _j)}$$
\vskip 0.5 cm
\noindent

Our goal is to reduce the model on the sub-manifold
of the phase space which does not contain the non-compact directions.
There is a general Dirac procedure to do that: we have to
fix $N-1$ first kind constraints (the integrals of motion $t_k^0$),
the non-compact coordinates being the ``auxiliary'' relations.
However, in the case under consideration 
there is a very direct way of describing the reduced
model which allows quantum analogue.  
To be precise, there is an $N\times N$ matrix $s$ whose matrix elements are
dynamical variables such that the similarity transformation of $l(z)$:
\begin{align}
m(z)=sl(z)s^{-1}\label{sls}
\end{align}
is of the form:
\begin{align}
&
m(z)=
\begin{pmatrix}
a(z) &b(z)
\non\\
c(z) &d(z) 
\end{pmatrix},
\end{align} 
where $d(z)$ is $(N-1)\times (N-1)$ matrix, $c(z)$ is $(N-1)$-vector,
$b(z)$ is $(N-1)$-covector and $a(z)$ is a scalar.
The similarity transformation (\ref{sls}) is such that 
$a(z)$ is a polynomial of degree $n-2$,
$c(z)$ is a polynomial of degree $n-1$,
$$b(z)=z^{n-1} e_{N-1}+\widetilde{b}(z),\quad d(z)=z^n d_0+\tilde{d}(z)$$
where $\widetilde{b}(z)$ is a polynomial of degree $n-2$,
$\tilde{d}(z)$ is a polynomial of degree $n-1$, 
the matrix elements of $d_0$ are in the center.
Here and later we use the notation:
\begin{align}
     &\begin{matrix}\  &\ &\ &\ \ j &\ &\  \end{matrix}\non\\
e_j= &\begin{pmatrix} 0 &\cdots &1 &\cdots &0\end{pmatrix}\non
\end{align}

Moreover, the coefficients of matrix elements of $m(z)$
have closed Poisson brackets. Thus the matrix $m(z)$ defines
the reduced 
phase space $\mathcal{M}_r$
and the reduced algebra of observables $\mathcal{A}_{r}$ as $l(z)$ defines 
$\mathcal{M}$
and $\mathcal{A}$. 
We do not write down corresponding Poisson brackets because
they can be obtained from quantum commutation relations described later.
In the classical case a similar construction for linear Poisson
brackets is given in \cite{ts}.

It is easy to calculate that
$$\text{dim}(\mathcal{M}_r)=2g$$
That is why
the algebra-geometric parametrization is well defined for the reduced model.
Namely, consider the symmetric power of the spectral curve:
$$X(g)=X^g/S_g$$
where $S_g$ is the symmetric group.
The points on $X(g)$ are divisors:
$P=\{p_1,\cdots ,p_g\}$ where $p_j=(z_j,w_j)$
are points on the spectral curve.
Consider the Abel transformation:
$$X(g)\ \to\ J$$
This transformation is one-to-one on non compact varieties:
$$X(g)-\widetilde{D}\simeq J_{aff}$$

With every $1\le i\le g$ we associate two numbers 
$k,l$ defined as 
follows.
$k$ is such that 
$$\frac 1 2 (k-1)(kn-2)<i\le\frac 1 2 k((k+1)n-2)$$
and $l$ is defined by
$$l=i-\frac 1 2 (k-1)(kn-2)$$
The polynomial $f_i(z,w)$ is defined by
\begin{align}
&f_i(z,w)=w^{k-1}z^{l-1}
\label{hd}
\end{align}
The meaning of $f_i(z,w)$ is clear: the holomorphic differentials
on the spectral curve are given by
$$\omega _i=\frac {f_i(z,w)dz}{\partial _w r(w,z)}
\qquad \forall i = 1,\ldots g$$

The divisor $\widetilde{D}$ is defined in terms of polynomials $f_i(z,w)$.
A point on $X(g)$ for which  $p_i\ne p_j\ \forall i\ne j$ belongs to
$\widetilde{D}$ if
$$\text{det}(f_i(z_j,w_j))=0$$
When there are coinciding points the definition needs some
changes, but we shall not go into details here.

There is an explicit construction:
$$m(z)\to X(g) $$
such that 
$$\{z _i,w_j\}=\delta _{i,j} z _iw_i$$
Thus the variables $z_j,w_j$ describe separated variables for the 
reduced integrable model. It is important that
the inverse map (algebraic) exists with
singularities on $\widetilde{D}$ only.
So, the level of integrals of the complexified reduced model 
give exactly the affine Jacobian. 

\section{Quantum reduction.}
Consider a quantum integrable model described by the l-operator
$L(z)$ satisfying standard commutation relations
\begin{align}
&R (z,z')(L(z)\otimes I)(I\otimes L(z'))
=(I\otimes L(z'))(L(z)\otimes I)R (z,z')
\label{RTT}
\end{align}
The quantum $R$-matrix is given by:
$$R (z,z')=zR_{12}(q)-z'R_{21}(q)^{-1}$$
where
$$q=e^{i\gamma},$$
$\gamma$ plays the role of Plank constant.
The constant R-matrix is given by
$$R_{12}(q)=\sum\limits _{j=1}^N q^{E^{jj}}\otimes q^{E^{jj}}
+(q-q^{-1})\sum\limits _{j>i}E^{ji}\otimes E^{ij}$$
Define the quantum determinant of $L(z)$:
$$\text{q-det}(L(z))=\sum\limits _{\pi}(-q)^{l(\pi)}
L_{1\pi (1)}(zq^{\frac {N-1} 2})
L_{2\pi (2)}(zq^{\frac {N-3} 2})\cdots L_{N\pi (N)}(zq^{-{\frac {N-1} 2}})
$$
where $l(\pi)$ is the minimal
number of transpositions of nearest neighbors in $\pi$.
\newline
Integrals of motion and elements of center appear in
$$\text{q-det}(wI+L(z))$$ 
Now we have to be a little bit more specific.
Like in classics (\ref{l}) we shall assume that 
the leading coefficient of $L(z)$ is lower-triangular:
$$L(z)=z^n\mu+O(z^{n-1}) $$
where
$$\mu =\begin{pmatrix} 
0        &0          &0                  &\cdots     &0     \non\\
\mu_{21} &\mu_{22}   &0                  &\cdots     &0       \non\\
\mu_{31} &\mu_{32}   &\mu_{33}           &\cdots     &0       \non\\
\vdots    & \vdots   & \vdots            &\vdots     &\ddots   \non\\
\mu_{N1} &\mu_{N2}   &\mu_{N3}           &\cdots     &\mu_{NN} 
\end{pmatrix}$$
such form is consistent with the commutation relations
(\ref{RTT}). Setting $\mu _{11}=0$ is very convenient for us, it corresponds to
some special choice of integrable model.

Take the first row of $L(z)$:
$$e_1L(z)=z^{n-1}\nu+O(z^{n-2}), $$
and
consider the $N\times N$ matrix whose matrix elements
are operators:
$$S=\begin{pmatrix} e_1\non\\ \nu\mu ^{N-2}\non\\\vdots \non\\
\nu\mu \non\\ \nu \end{pmatrix}$$
\noindent
Let us show that 
\begin{align}
&S\mu=US\non
\end{align}
where
\begin{align}
&U=\begin{pmatrix} 
0 &0              &\cdots   &0                &0           \non\\
0 &-t_1        &\cdots   &-t_{N-2}         &-t_{N-1}     \non\\
0 &1            &\cdots   &0                &0           \non\\
\vdots       & \vdots  &\ddots           &\vdots      \non\\
0 &0               &\cdots   &1 \vdots &0
\end{pmatrix},
\end{align}
and $t_j$  are defined by
$$\prod\limits _{i=2}^N (x-q^{-1}\mu _{jj})=\sum\limits _{j=0}^{N-1}
x^{N-1-j}t_j$$
Obviously, the only thing we have to prove is 
\begin{align}
& \sum\limits _{k=0}^{N-1}t_{k}\nu\mu ^{N-1-k}=0\label{ch}
\end{align}
where $t_0=1$. Notice two circumstances.
First, the relations (\ref{RTT}) imply that
\begin{align}
& \mu _{jj}q^{E^{jj}}\mu =\mu q^{E^{jj}}\mu _{jj}\non\\
&\nu\mu_{jj}=\mu _{jj}\nu\label{xx}
\end{align}
Second, the expression $\mu ^k$ contains 
\begin{align}
&\mu _{j_1j_2}\mu _{j_2j_3}\mu _{j_3j_4}\cdots
\label{mmm}
\end{align}
with $j_1\ge j_2\ge j_3\ge j_4\ge\cdots$. From (\ref{xx})
one finds:
\begin{align}
&\nu\mu ^l\mu _{jj}E^{jj}=\nu \mu ^l\mu _{jj}q^{E^{jj}-1}E^{jj}=
q^{-1}\mu _{jj}\nu\mu ^lE^{jj}
\non
\end{align}
So, in the expression $\mu ^k$ one can  move to
the left all the $\mu _{jj}$  replacing them
by $q^{-1}\mu _{jj}$, the rest consists of 
expressions of the form (\ref{mmm}) with strictly ordered indices.
Hence the equation (\ref{ch}) follows from the corresponding classical
equation: a matrix with commuting entries satisfies its characteristic
equation.

Introduce the matrix $M(z)$:
\begin{align}
M(z)=SL(z)S^{-1}
\non
\end{align}
We have shown that 
\begin{align}
M(z)=z^nU+O(z^{n-1})
\non
\end{align}
Moreover, since $\nu $ is the last row of $S$ we have 
$$e_1M(z)=z^{n-1}e_N+O(z^{n-2})$$
It is important that $t_j$ commute with elements of $M(z)$.

Similarly to the classical case introduce the notations
\begin{align}
&
M(z)=
\begin{pmatrix}
A(z) &B(z)
\non\\
C(z) &D(z) 
\end{pmatrix},
\non
\end{align} 
where $D(z)$ is $(N-1)\times (N-1)$ matrix etc.
We have the same behavior in $z$ as in classics:
$A(z)$ is a polynomial of degree $n-2$,
$C(z)$ is a polynomial of degree $n-1$,
$$B(z)=z^{n-1}e_{N-1}+\widetilde{B}(z),\quad
D(z)=\widetilde{U}z^n+\widetilde{D}(z)$$
where $\widetilde{B}(z)$, $\widetilde{D}(z) $ are respectively
polynomials of degrees $n-2$ and $n-1$, $\widetilde{U}$ is $(N-1)\times (N-1)$
matrix obtained from $U$ by omitting first row and first column.

The algebra $\mathcal{A}_q$ is generated by 
coefficients of matrix elements of $M (z)$,
$t_j$ can be considered as
$\mathbb{C}$-number parameters of $\mathcal{A}_q$.

We wand to find 
closed commutation relations for $M(z)$.
Introduce the following object:
\begin{align}
&\widehat{S}_{12}=qE^{11}\otimes q^{E^{11}}+
\sum\limits _{i=2}^N
(\nu\otimes I)(R_{12}(q)(\mu\otimes I))^{N-i}(E^{ii}\otimes I)\non
\end{align}
which is a matrix in the tensor product $\mathbb{C}^N\otimes \mathbb{C}^N$
with non-commuting entries. 
It is possible to prove the 
auxiliary commutation relations:
\begin{align}
&Y_{12}(S\otimes I)\widehat{S}_{21}=(I\otimes S)\widehat{S}_{12}R_{12}(q),
\non\\
&(S\otimes I)\widehat{S}_{21}(T(z)\otimes I)
=Z_{12}(z)
(M(z)\otimes I)(S\otimes I)\widehat{S}_{21}R_{21}(q)
\label{au}
\end{align}
where we use the usual notation $X_{21}=PX_{12}P$ with $P$ being the permutation
matrix acting in $\mathbb{C}^N\otimes \mathbb{C}^N$.
The matrices $Y_{12}$ and $Z_{12}(z)$ have $\mathbb{C}$-number matrix elements.
They are defined as follows.
Consider the matrix
$$C_{12}=(I-E^{11})\otimes I+\sum\limits _{j=1}^{\infty}V^j\otimes U^j$$
where
\begin{align}
&V=\begin{pmatrix} 
0      &0         &0                    &\cdots     &0       \non\\
0      &0         &1                   &\cdots     &0       \non\\
\vdots &\vdots   & \ddots       &\vdots     &\vdots  \non\\
0 &0   &0         &0               &1       \non\\
0 &0   &0         &0               &0
\end{pmatrix}
\non
\end{align}
The matrices $Y_{12}$ and $Z_{12}(z)$ are defined using $C_{12}$:
\begin{align}
&Y_{12}=qI-(q-q^{-1})C_{12}(I-P),\non\\
&Z_{12}(z)=I-(q-q^{-1})z\ (I\otimes U)C_{12}(E^{N,1}\otimes I)\non
\end{align}
where $P$ is permutation.

It is easy to see that the operator $C_{12}(I-P)$ is a projector:
$$\(C_{12}(I-P)\)^2=C_{12}(I-P)$$
which explains the following simple formula for the inverse of $Y_{12}$:
$$Y_{12}^{-1}=q^{-1}I+(q-q^{-1})C_{12}(I-P)$$

Now we are ready to find the closed commutation relations in question.
Take the relation (\ref{RTT}) and
multiply it by 
$(I\otimes S)\widehat{S}_{12}$
from the left 
and by $((S\otimes I)\widehat{S}_{21})^{-1}$ from
the right. After some calculations using (\ref{au}) one finds
\begin{align}
&\widetilde{R}(z_1,z_2)K_{12}(z_1)(M(z_1)\otimes I)
K_{21}(z_2)(I\otimes M(z_2))=\non\\
&=K_{21}(z_2)(I\otimes M(z_2))
K_{12}(z_1)(M(z_1)\otimes I)\widetilde{R}(z_1,z_2)\non
\end{align}   
where
\begin{align}
&\widetilde{R}(z_1,z_2)=z_1Y_{12}-z_2Y_{21}^{-1},\non\\&
K_{12}(z)=Y_{12}^{-1}Z_{12}(z)\non
\end{align} 

\section{Discussion.}

Similarly to \cite{skl,skl2,sepN} 
it is possible to construct quantum separated variables:
$$M(z)\to (z_1, w_1),\cdots ,(z_g,w_g)$$
such that
$$w_iz_i=q^2 z_iw_i$$
It should be possible to show that 
every element of $\mathcal{A}_q$ can be expressed in terms of 
these separated variables. This is a complicated statement which we
did not prove yet, but exactly this property of $\mathcal{A}_q$
is the reason for introducing reduced model. An analogue
of singularities on $\widetilde{D}$ that we had
in classical has to manifest itself after quantization.

The original Hilbert space where the elements of $\mathcal{A}_q$
act should be unitary equivalent to 
the Hilbert space: functions of 
$$\zeta _j=\frac 1 2 \text{log}(z_j)$$
The operator $w_j$ is shift of $\zeta _j$ by $i\gamma$.
To finish the definition of the Hilbert space we need
to define the scalar product. 
The duality of quantum integrable model is to
be emphasized.
There is dual integrable model which is similar to
the original one but for which the Plank constant
changes
$$\gamma\ \to\ \frac {\pi ^2}{\gamma}$$  
Consider the dual operators 
$$Z_j=\exp \bigl(\frac {2\pi}{\gamma}\zeta _j\bigr),\ 
W_j=\exp \bigl(\pi i\frac 
{\partial}{\partial\zeta _j}\bigr)$$
It is easy to see that formally $Z_j$, $W_j$ commute with
$z_j$, $w_j$. But, again, to make real sense of the commutativity
we have to define the Hilbert space where the operators act.
We would conjecture that the scalar product in the space
of functions of $\zeta _j$ is given by
\begin{align}
\langle F\ |\ G\rangle =&\int\limits _{-\infty}^{\infty} d\zeta _1
\cdots \int\limits _{-\infty}^{\infty} d\zeta _g  
\ \overline{ F(\zeta _1,\cdots ,\zeta _g )}\non\\
&\times\text{det}(f_i(z_j,w_j))\text{det}(f_i(Z_j,W_j))
G(\zeta _1,\cdots ,\zeta _g )\non
\end{align}
where $f_i$ are the polynomials defining the holomorphic differentials
(\ref{hd}). Notice that $w_j$, $W_j$ act on $G(\zeta _1,\cdots ,\zeta _g )$.
There are two reasons why we believe this formula to be true.
First of them is quasiclassics:
the determinant $\text{det}(f_i(z_j,w_j))$ enters
Liouville measure rewritten in the separated variables. 
The second is known case $N=2$ \cite{qaf}.
We shall comment more on this point in later publications.

\end{document}